\documentclass[superscriptaddress]{revtex4}

\usepackage{verbatim}
\usepackage{graphicx}
\usepackage{multirow}
\usepackage{subfigure}
\usepackage{amsfonts}
\usepackage{amsmath}
\usepackage{amsbsy}
\usepackage{color}
\usepackage[colorlinks=true, urlcolor=blue, citecolor=blue, linkcolor=blue]{hyperref}
\usepackage{bm}
\usepackage{esint}

\let\sb=_ \catcode`\_=\active \def_#1{\ensuremath \sb{\rm#1}}

\begin{document}

\title{Domino-like magnetic phase transition induced by a bias voltage in FeRh thin film}

\author{Huiliang Wu}
\affiliation{School of physical science and technology, Lanzhou University, Lanzhou 730000, People's Republic of China}

\author{Jianbo Wang}
\email{wangjb@lzu.edu.cn}
\affiliation{School of physical science and technology, Lanzhou University, Lanzhou 730000, People's Republic of China}

\author{Chenbo Zhao}
\affiliation{School of physical science and technology, Lanzhou University, Lanzhou 730000, People's Republic of China}

\author{Qingfeng Zhan}
\affiliation{Key Laboratory of Polar Materials and Devices (MOE), School of Physics and Electronic Science, East China Normal University, Shanghai 200241, People's Republic of China}

\author{Jiangtao Xue}
\affiliation{School of physical science and technology, Lanzhou University, Lanzhou 730000, People's Republic of China}

\author{Senfu Zhang}
\affiliation{School of physical science and technology, Lanzhou University, Lanzhou 730000, People's Republic of China}

\author{Jinwu Wei}
\affiliation{School of physical science and technology, Lanzhou University, Lanzhou 730000, People's Republic of China}

\author{Xiangqian Wang}
\affiliation{Key Laboratory of Sensor and Sensor Technology, Institute of Sensor Technology, Gansu Academy of Science, Lanzhou, 730000, People's Republic of China}

\author{Qingfang Liu}
\email{liuqf@lzu.edu.cn}
\affiliation{School of physical science and technology, Lanzhou University, Lanzhou 730000, People's Republic of China}

\begin{abstract}
The first-order magnetic phase transition (MPT) usually happens with a very wide magnetic field span about tens of thousands Oersted which hinders its applications. In this work, we induce a domino-like MPT via introducing a bias voltage in FeRh thin film and thus realize a large narrowing of transition magnetic field span from $6\times10^4$ Oe to lower than $2\times10^3$ Oe at room temperature. Meanwhile, nonvolatile switchings between pure magnetic phases of FeRh at room temperature also can be realized by applying voltage pulses. The critical condition and phase diagram for domino-like MPTs are obtained in theory and our experiments support it well. Our works not only benefit the studies and applications of MPT-based devices but also are significant in the applications of the phase transition systems with resistance change.
\end{abstract}

\maketitle

\section{Introduction}
The first-order magnetic phase transition (MPT) possesses the transition in the spin order degree of freedom and the characteristic properties of the first-order phase transition such as volume change, entropy change, latent heat, and the existence of metastable phase et.al. \cite{1}. In addition to the common natures of the first-order MPT mentioned above, there are various materials occurring the first-order MPT accompanied with their special properties, such as the giant magneto-caloric effect \cite{2,3}, the metal-insulator transition \cite{4,5}, and the shape memory \cite{6,7}, et.al. Profited from the multi-physics properties in various first-order MPT materials, the first-order MPTs provide important platforms for researching fundamental physics \cite{7,8,9,10,11} and are significant in many applications in magnetic refrigeration \cite{11,3,12}, memory resistors \cite{13,14,15,16}, magnetostriction materials \cite{20,21}, and spintronic devices \cite{17,10,18,19}, et.al.

The magnetic field is a special and effective driving force \cite{2,3,4,5} for the first-order MPTs because the Zeeman energy is involved in the free energies of magnetic phases. Thus, phase transition induced by a magnetic field in MPT materials is an important property in the applications mentioned above. However, the transition span of the magnetic field, namely the span of the magnetic field from the start to the end of MPTs, usually reach about tens of thousands Oersted \cite{2,3,5,11,12}. Such a large magnetic field only can be supplied by a superconducting magnet, which greatly hinders the applications of MPTs. In addition, the large mixed-phases region in the MPT loop caused by the broadened transition spans makes the nonvolatile switchings between pure phases in MPT materials difficult to occur, which is not conducive to achieving a large change ratio of physical properties such as the resistance \cite{16}, the magnetic damping \cite{34}, and the spin-orbit coupling \cite{35} et al. in MPT-based devices.

The broadened transition spans result from the distribution of transition points due to the disorders and inhomogeneity in materials \cite{1,33}, which also has been described phenomenologically \cite{22,16,23} using the Preisach model in MPT systems. In order to narrow the transition span, researchers have tried to reduce the temperature to below 5 K to suppress the thermal-assisted nucleation in MPTs \cite{24,25}, and fabricate the materials into mesoscale (0.55 $\mu$m) to reduce the amount of effective nucleation site in ferromagnetic-antiferromagnetic (FM-AFM) transitions \cite{26}. Although suppressing the nucleation can lead to a steep MPT, the requirement of the extremely low temperature or small scale for effectively suppressing nucleation still has significant limitations in practical applications. How to effectively reduce the transition span in practical applications remains an open problem.

CsCl-type FeRh is an important MPT alloy possessing first-order MPT between AFM and FM phases just above room-temperature (about 370 K), which makes it a convenient test bed for exploring the fundamental physics and developing the applications of MPTs \cite{10,13,30} compared with other materials possessing first-order MPTs with temperature far below room-temperature \cite{1}. The first-order MPTs are usually accompanied by large changes in electrical resistance. In this work, we demonstrate a voltage-induced chain-like FeRh MPT just like the domino in experiment and theory. By measuring the resistance-temperature (R-T) loops under various bias voltages and resistance-voltage (R-V) loops under different temperatures, and analyzing the thermal equilibrium equations, we find the domino-like transition is caused by the chain-like changes of resistance during the first-order MPTs under a certain bias voltage. As a result, (1) nonvolatile switchings between pure magnetic phases of FeRh at room temperature are realized by applying voltage pulses; (2) a magnetic-field-driven MPT with a greatly narrowed transition span is easily realized by applying a bias voltage and a tabletop electromagnet. In the end, the critical condition and the phase diagram for domino-like MPT are summarized in theory, which corresponds well with our measurements.

\section{Results and discussion}

\subsection{Experiment setup and FeRh MPT under bias voltages}

\begin{figure}[h]
\centering
\includegraphics[width=0.8\textwidth]{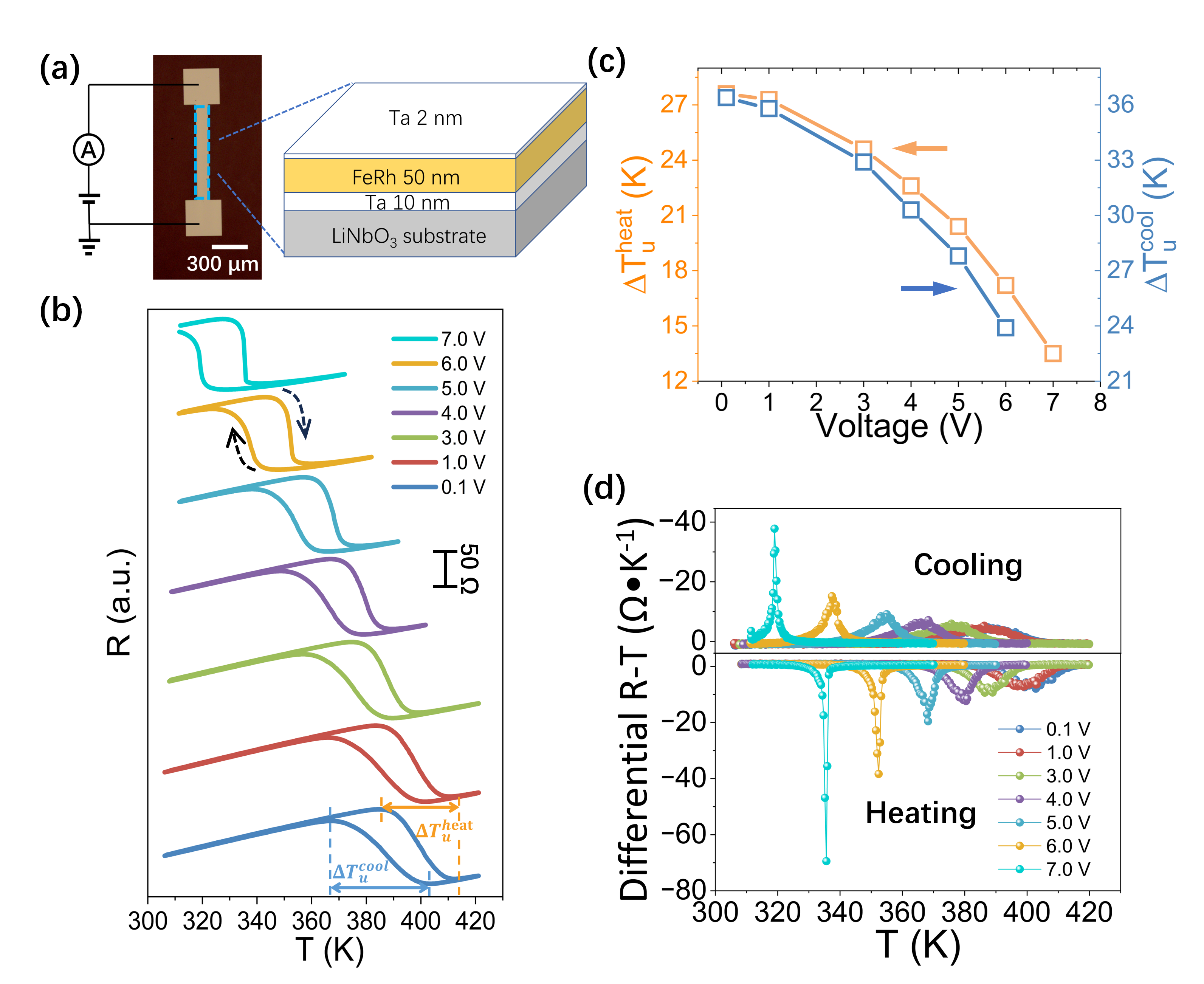}
\caption{Experiment setup and FeRh MPT under various bias voltages. (a) The illustration of our device and the schematic diagram of the film structure. (b) The R-T loops of FeRh under different bias voltages. (c) The transition span extracted from the R-T loops under different bias voltages in (b). (d) The differential R-T curves under different bias voltages transformed from the corresponding R-T loops in (b).}
\label{1}
\end{figure}

We apply series bias voltages to the FeRh strip (figure \ref{1} (a)) and then measure the R-T loops under different bias voltages. There is almost no difference between the results of 1.0 V and 0.1 V, thus the 0.1 V bias voltage only works as a probe voltage to measure the resistance. The results are shown in figure \ref{1} (b). The R-T loops have steeper transitions in both the heating and cooling transition branches as the bias voltage increases. Meanwhile, the R-T loop shifts to the lower temperature region as the bias voltage increases. We define the temperature difference between the points with zero slop in a R-T branch as the transition span ($\Delta T_u^{heat}$ and $\Delta T_u^{cool}$), as shown in figure \ref{1} (b). The $\Delta T_u^{cool}$ and $\Delta T_u^{heat}$ as a function of the bias voltage is shown in figure \ref{1} (c). The differential R-T curves shown in figure \ref{1} (d) shows that the minimum slop of the heating branch changes from about -7 $\Omega$K$^{-1}$ to about -70 $\Omega$K$^{-1}$, while the cooling branch changes from about -5 $\Omega$K$^{-1}$ to about -38 $\Omega$K$^{-1}$. It means that the bias voltage significantly improves the phase transition. Similar results are observed in the FeRh strip sputtered on the MgO substrate (in the supporting information).

\subsection{Bias voltage-driven MPT and the transformation of R-V loops}

\begin{figure}
\centering
\includegraphics[width=1\textwidth]{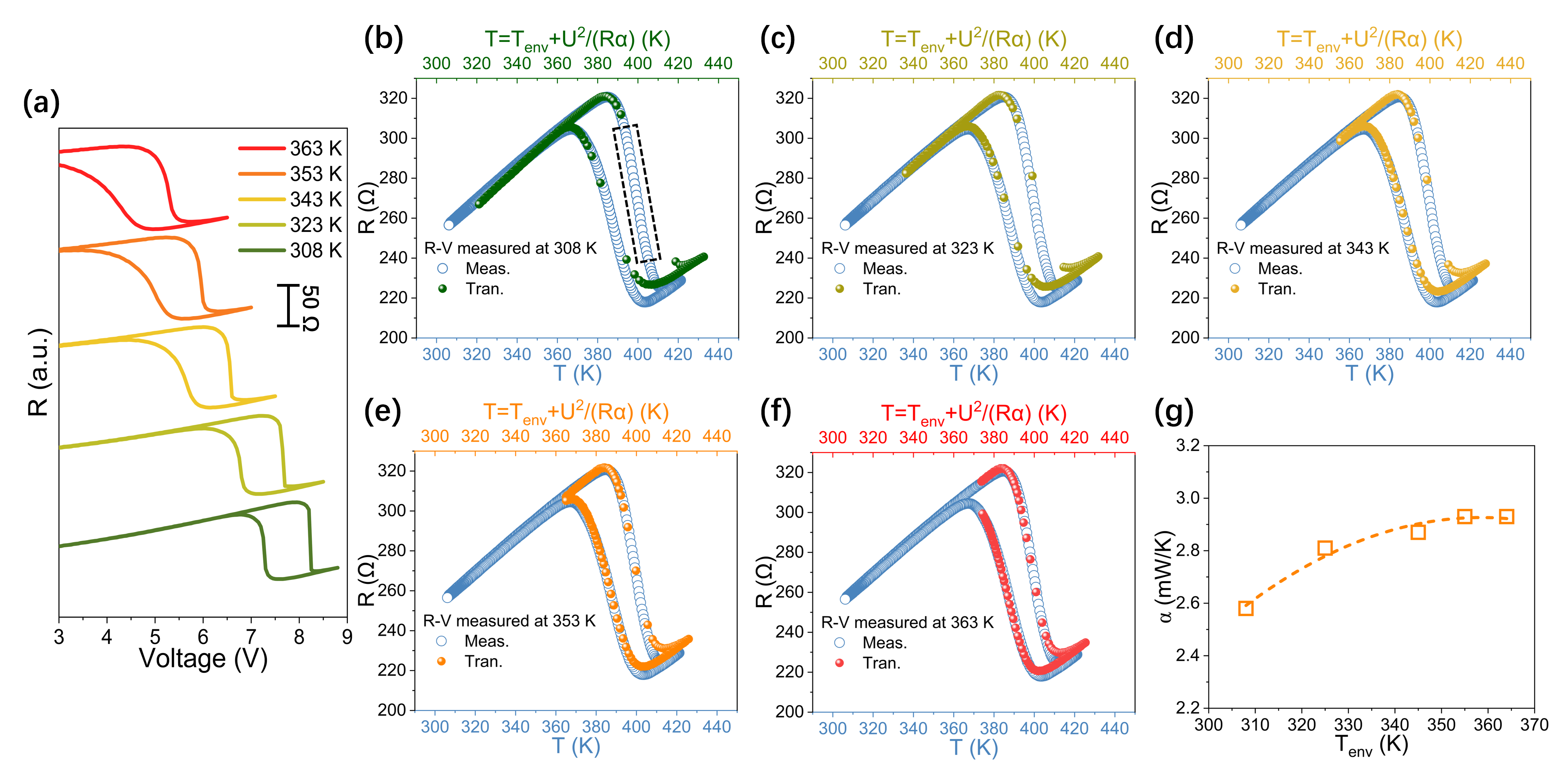}
\caption{The voltage-driven MPT and the transformation from R-V loops to R-T loops. (a) The voltage-induced FeRh MPT loops (R-V loops) under different environment temperatures. (b)-(f) The transformed R-T loops (marked as Tran.) from R-V loops in (a). The measured R-T loop with 0.1 V probe voltage is also displayed for comparison with the transformed one. (g) The coefficients of the heat dissipation under different environment temperatures.}
\label{2}
\end{figure}

Figure \ref{2} (a) shows that the transition happens at higher voltages as the temperature decreases, and the transition loops are steeper at higher voltages. Noriko Matsuzaki et al. have proposed a transformation formula $\alpha(T-T_{env})=I^2R$, where $\alpha$ is the coefficient of heat dissipation and $T_{env}$ is the environment temperature, to transform the R-current loop to a R-T loop \cite{27}. This formula indicates the power of the heat dissipation equal to the power of the Joule heating of electric current when the sample reaches thermal equilibrium. Based on the same principle, we have the transformation formula $\alpha(T-T_{env})=U^2/R$ to transform R-V loops to R-T loops (figure \ref{2} (b)-(f)). These transformed loops correspond well with the measured R-T loop, especially for R-V loops with MPT occurring at lower voltages (figure \ref {2} (e) and (f)). In figure \ref{2} (b), there is no data point during the main heating transition (marked in the dashed black rectangle) for the transformed loop. It is because, at higher voltages (about 8 V), the transition (see the heating transition of the 308 K in figure \ref {2} (a)) is too steep thus no data point can be captured in experiment. In the discussion part, we further know there is a domino-like MPT thus we can not measure any data point in that region. In addition, in figure \ref{2} (b)-(e), there is some mismatch between the measured and transformed results at FM state. It is because the transformation formula treats the sample with homogenetic temperature, but in fact, the temperature difference between the edges and the center of the strip exists when a larger voltage is applied. It also explains why the transformed loop match better when the transition occurs at lower voltages (figure \ref{2} (e) and (f)). The coefficient of heat dissipation $\alpha$ is determined by satisfying the transformed loop matching well with the measured loop. The results of $\alpha$ changing with environment temperature are displayed in figure \ref{2} (g). The $\alpha$ at 308 K is 2.57 mW/K, which is close to the result of $\alpha=2.43$ mW/K at 303 K in our previous work \cite{16}.

\subsection{Bias voltage-induced domino effect and the calculated MPT loops}

When the FeRh strip is in the environment with environment temperature $T_{env}$ and is applied a certain bias voltage $U$, the temperature of FeRh ($T_{sam}$) will be higher than the environment temperature due to the Joule heating. In this case, we have
\begin{equation}
    \frac{U^2}{R}=\alpha(T_{sam}-T_{env})
\end{equation}
to describe thermal equilibrium. Next, if $T_{env}$ changes $dT_{env}$, to rebuild the thermal equilibrium, the resistance of FeRh will change $dR$ and $T_{sam}$ will changes $dT_{sam}$. Thus, in this new thermal equilibrium state, we have
\begin{equation}
    \frac{U^2}{R+dR}=\alpha[(T_{sam}+dT_{sam})-(T_{env}+dT_{env})].
\end{equation}
Combined the equation (1)(2), we have
\begin{equation}
    (\frac{\partial R}{\partial T_{env}})^{-1}=(\frac{\partial R}{\partial T_{sam}})^{-1}+\frac{U^2}{R^2\alpha}.
\end{equation}
$(\partial R/\partial T_{sam})^{-1}$ can be gained from the differential R-T curve with 0.1 V probe voltage in figure \ref{1} (d), and $\alpha$ changes with $T_{env}$ has been summarized in figure \ref{2} (g).

\begin{figure}
\centering
\includegraphics[width=0.9\textwidth]{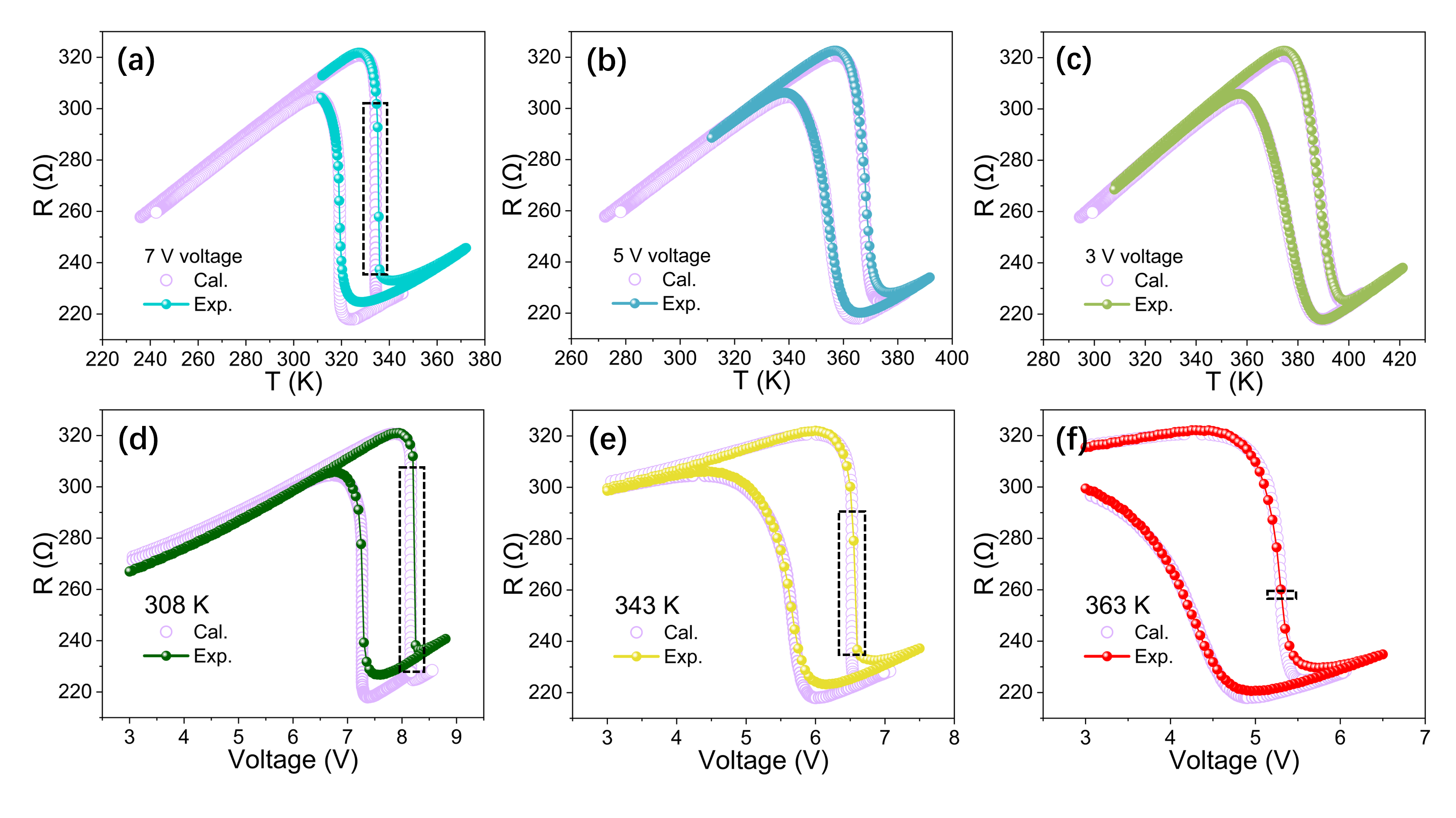}
\caption{The calculated MPT loops compared with the measured one. (a)-(c) The calculated R-T loops (marked as Cal.) and the measured R-T loops (marked as Exp.) under different bias voltages. (d)-(f) The calculated R-V loops and the measured R-V loops under different environment temperatures.}
\label{3}
\end{figure}

Using equation (3), we calculate the loops under various bias voltages (figure \ref{4} (a)-(c) and figure S2 in supporting information). The calculated results correspond well with our measurements, meaning the equation (3) can be used to describe the phenomena of temperature-induced FeRh MPT under the bias voltages. In the calculation of the 7 V voltage (figure \ref{3} (a)), a vertical drop of resistance (marked in the black dashed rectangle) in the heating transition branch is observed, while in the cooling transition branch, although the vertical switching of resistance also exists, but the extent is smaller than that in the heating branch. These steep changes of resistance indicate that a tiny perturbation is enough to trigger the phase transition. Return back to equation (1), in qualitative, we can know there should be a chain change: if $T_{env}$ increases during the heating transition, $R$ will decrease, then it causes increases in the Joule heating power $U^2/R$ and sample's temperature $T_{sam}$, resulting in a further decrease of $R$... again and again, just like a domino effect. In quantitative, because the slop $\partial R/\partial T_{env}$ must be negative during the MPT, we obtain that the domino-like transition will occur when
\begin{equation}
    (\frac{\partial R}{\partial T_{sam}})^{-1}+\frac{U^2}{R^2\alpha}\geq0.
\end{equation}

In the case of triggering MPT by changing the voltage, if $U$ changes $dU$, to rebuild the thermal equilibrium, the resistance of FeRh will change $dR$ and $T_{sam}$ will change $dT_{sam}$. Thus, we have
\begin{equation}
    \frac{(U+dU)^2}{R+dR}=\alpha[(T_{sam}+dT_{sam})-T_{env}].
\end{equation}
Combined equation (1)(5), we gain
\begin{equation}
    (\frac{\partial R}{\partial U})^{-1}=\frac{R\alpha}{2U}(\frac{\partial R}{\partial T_{sam}})^{-1}+\frac{U}{2R}.
\end{equation}
Using equation (6), we calculate the R-V loops under the different environment temperatures (figure \ref{3} (d)-(f) and figure S2 in supporting information). In figure \ref{3} (d), we can see a more obvious vertical drop of resistance, namely a domino-like MPT, in both calculated and experiment results, because the higher voltage is more easily to induce the domino-like MPT. In the calculated loops in figure \ref{3} (d)-(f), we mark the vertical drops of resistance in black dashed rectangles indicating the domino-like MPT. From equation (6), because the slop $\partial R/\partial U$ must be negative during the MPT, we also can derive an identical inequation to the inequation (4) to judge whether the domino-like transition will occur in the case of triggering MPT by changing the bias voltage.

\subsection{Voltage-pulse-triggered nonvolatile switching between pure magnetic phases in FeRh}

\begin{figure}[h]
\centering
\includegraphics[width=0.7 \textwidth]{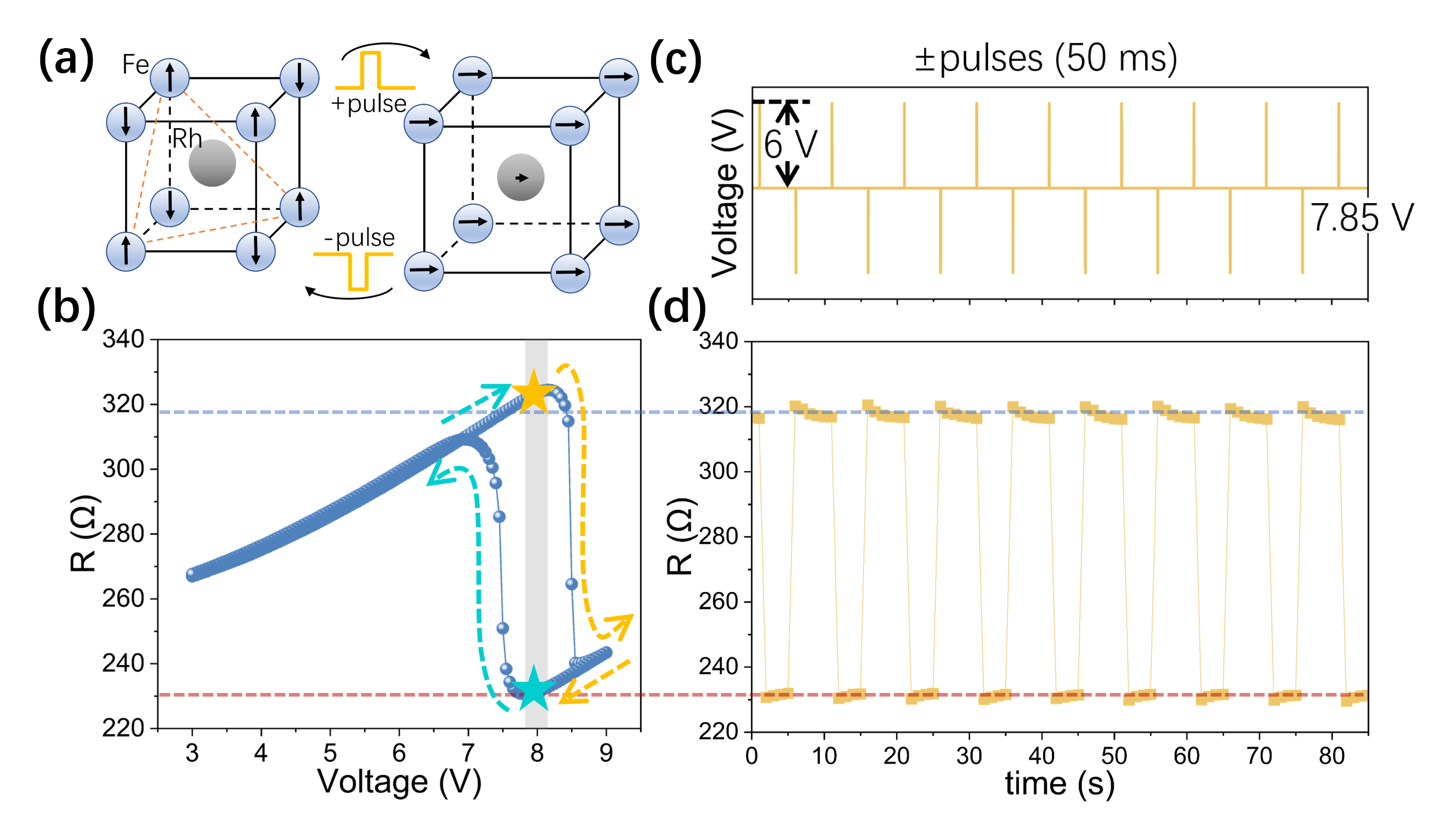}
\caption{Nonvolatile switching between pure magnetic phases via voltage pulses. (a) The schematic diagram of the magnetic phases switching via positive and negative pulses. (b) R-V loop at room temperature. The region marked by a gray shadow allows the existence of both the pure AFM and the pure FM phases. (c) Positive (+) and negative (-) pulses with 50 ms pulse width were applied in this experiment. The +pulses and -pulses are superposed on a 7.85 V voltage platform. (d) Repeated nonvolatile switching between pure AFM and pure FM phases triggered by $\pm$pulses superposed on a 7.85 V voltage platform at room temperature.}
\label{4}
\end{figure}

Based on the results in figure \ref{2}, once the domino-like MPT happens, nonvolatile switchings between pure AFM and pure FM phase at room temperature can be realized by applying voltage pulses, just like the schematic diagram in figure \ref{4} (a). We set a 7.85 V locating within the gray shadow in figure \ref{4} (b) as the constant voltage platform, and then superpose a positive voltage pulse (+pulse) on the constant voltage platform to trigger the FeRh transition from the pure AFM phase to the pure FM phase. The orange dashed arrows in figure \ref{4} (b) are the schematic diagram of the MPT path when a +pulse is applied at the orange star, which means the FeRh strip will be heated transiently to the FM state within the pulse's duration and then rapidly cool down to the state at the cyan star after the end of the pulse. To return to the pure AFM state, we apply a negative voltage pulse (-pulse) superposed on the constant voltage platform. The cyan dashed arrows in figure \ref{4} (b) show the MPT path when a -pulse is applied at the cyan star, implying the FeRh strip will be cooled transiently to AFM state within the pulse's duration and then rapidly heat up to the state at the orange star after the end of the pulse. The nonvolatile switching results are in figure \ref{4} (d). The +pulse switches the FeRh into low resistance and the FeRh keeps at a low resistance state after the +pulse is applied. The rad dashed guide line between figure \ref{4} (b) and (d) implies that the nonvolatile low resistance state is a pure FM state. The FeRh states with high resistances after the -pulse also are nonvolatile. The blue dashed guideline implies that the nonvolatile high resistance state is nearly a pure AFM state.

\subsection{Magnetic-field-driven MPT under the bias voltages}

\begin{figure}[h]
\centering
\includegraphics[width=1 \textwidth]{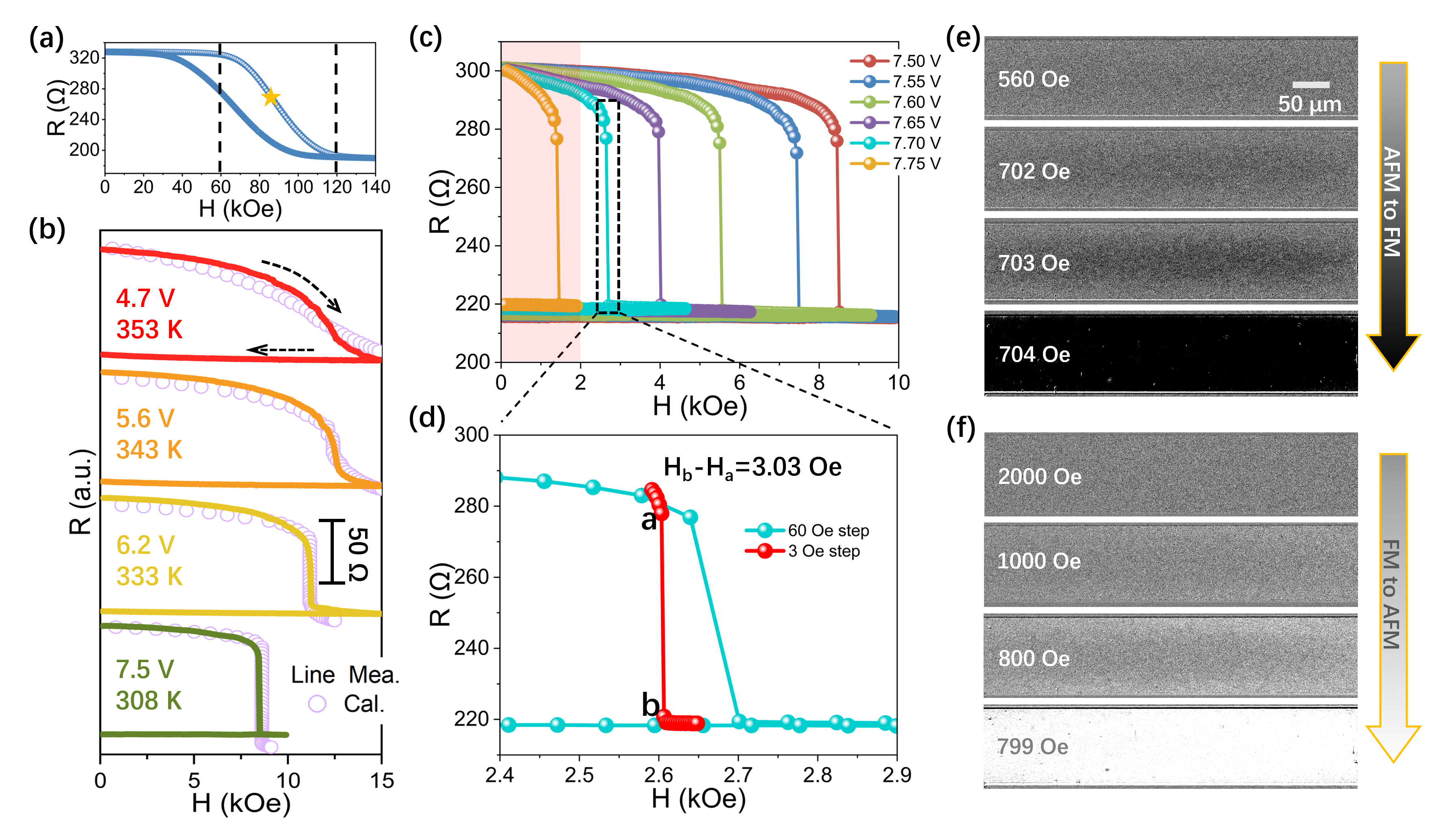}
\caption{Magnetic-field-driven FeRh MPT under the voltage. (a) The R-H loop without applied the bias voltage. (b) The AFM-FM transition curves under different voltages compared with the calculated results. Different environment temperatures are used to shift the AFM-FM transition under various bias voltages to the magnetic field between 0 and 15 kOe. (c) The AFM-FM transition curves tuned via adjusting the voltage near 7.7 V at room temperature. The magnetic field step is about 60 Oe. (d) Finely measured AFM-FM transition compared with the enlarged result of 7.7 V in (c). (e) Images of magnetic phase domain evolution induced by the magnetic field under 7.78 V voltage at AFM-FM transition. (f) Images of magnetic phase domain evolution induced by the magnetic field under 7.00 V voltage at FM-AFM transition.}
\label{5}
\end{figure}

As another important application, our discoveries are significant in the magnetic-field-driven MPTs, because the process of MPT in FeRh and other MPT materials universally needs magnetic field with values of several ten thousand Oersted when using a magnetic field to drive MPT \cite{2,3,5,11,12}. Figure \ref{5} (a) is the R-H loop only with a 0.1 mA probe current applied. The transition span of AFM-FM transition is about $6\times10^4$ Oe (the extent of the magnetic field between two black dashed lines). The steepest slope of the AFM-FM transition is $-3.59\times10^{-3}$ $\Omega\cdot$Oe$^{-1}$ at 86 kOe (marked by a star). As the bias voltage is applied (figure \ref{5} (b)), the AFM-FM transition curve changes steeper with the voltage increasing, which demonstrates an electrically tunable transition span in magnetic field. Note that the environment temperature only shift the MPT point with negligible influence on the shape of the R-H loop \cite{32}. At 7.5 V and 6.2 V (the green and the yellow curves in figure \ref{5} (b)), there are sudden drops of R at the magnetic field near 8.5 kOe and 11.1 kOe, respectively. In these two voltages, the domino-like MPT happens and will be discussed later (which can be seen in the phase diagram in figure \ref{6}). Even at 4.7 V (the red curve), although the sudden transition vanishes, the AFM-FM transition still mainly occurs within 15 kOe.

We adjust the voltage finely to shift the AFM-FM transition (figure \ref{5} (c)). All of the R-H curves have a sudden drop of R and the transition point shifts from 8.5 kOe to 1.4 kOe with the voltage adjusted from 7.50 V to 7.75 V. At 7.75 V, the MPT magnetic field is less than 2 kOe (marked with a transparent red shadow), which is easily generated by most of the tabletop electromagnets. In figure \ref {5} (d), there is still a sudden drop of R even when 3 Oe magnetic field step is used to change the magnetic field. The magnetic field difference between the contiguous measurement points (a and b) is only 3.03 Oe. The slop of this steep transition reaches $-18.5$ $\Omega\cdot$Oe$^{-1}$, which is enhances by $10^4$ times compared with $-3.59\times10^{-3}$ $\Omega\cdot$Oe$^{-1}$ without applied bias voltage in figure \ref{5} (a). The FM-AFM transition (in the supporting information) also has similar phenomena as the AFM-FM transition discussed here.

Furthermore, figure \ref{5} (e) is the magnetic phase domain (MPD) images of the AFM-FM transition induced by a magnetic field under a 7.78 V voltage. When the magnetic field doesn't exceed 703 Oe, there are only some tiny black domains (FM phase domains). Once the magnetic field increases 1 Oe (the smallest step of our tabletop electromagnet) from 703 Oe, the contrast of the strip suddenly changes to black, which implies FeRh occurs a complete AFM-FM transition within only 1 Oe. The FM-AFM transition is similar to the AFM-FM transition, which can be seen in figure \ref{5} (f). When the magnetic field decreases 1 Oe from 800 Oe, the contrast of the strip suddenly changes to white, which implies FeRh also can occur a complete FM-AFM transition within only 1 Oe. The above results show that even a small magnetic field can drive the MPT once the domino-like MPT is introduced via applying a bias voltage.

The magnetic field represents another important type of driving force. This type of driving force, here we use Y to represent them, includes all of driving forces besides environment temperature and voltage, such as magnetic field, static strain, dynamic strain, et.al. In this case, if the driving force $Y$ changes $dY$, to rebuild the thermal equilibrium, the resistance of FeRh will change $dR$ and $T_{sam}$ will change $dT_{sam}$. Thus, we have
\begin{equation}
    \frac{U^2}{R+dR}=\alpha[(T_{sam}+dT_{sam})-T_{env}],
\end{equation}
and
\begin{equation}
    dR=\frac{\partial R}{\partial T_{sam}}dT_{sam}+\frac{\partial R}{\partial Y_0}dY.
\end{equation}
The $\partial R/\partial Y_0$ is the slop of the R-Y loop without bias voltage applied (figure \ref{5} (a)). Combined equation (1)(7)(8), we gain
\begin{equation}
    (\frac{\partial R}{\partial Y})^{-1}=[\frac{U^2}{R^2\alpha}(\frac{\partial R}{\partial T_{sam}})+1](\frac{\partial R}{\partial Y_0})^{-1}.
\end{equation}
Based on the equation (9), we calculate the R-H curves under different bias voltages and different environment temperatures (figure \ref{5} (b)). Moreover, although equation (9) is completely different from equation (3) and (6), considering the slop $\partial R/\partial Y$ must be negative during MPT, we also can gain an identical inequation to the inequation (4) from equation (9) to judge whether the domino-like transition will occur in the case of driving force $Y$.

\subsection{The critical condition and the phase diagram for domino-like MPTs}

\begin{figure}[h]
\centering
\includegraphics[width=0.8\textwidth]{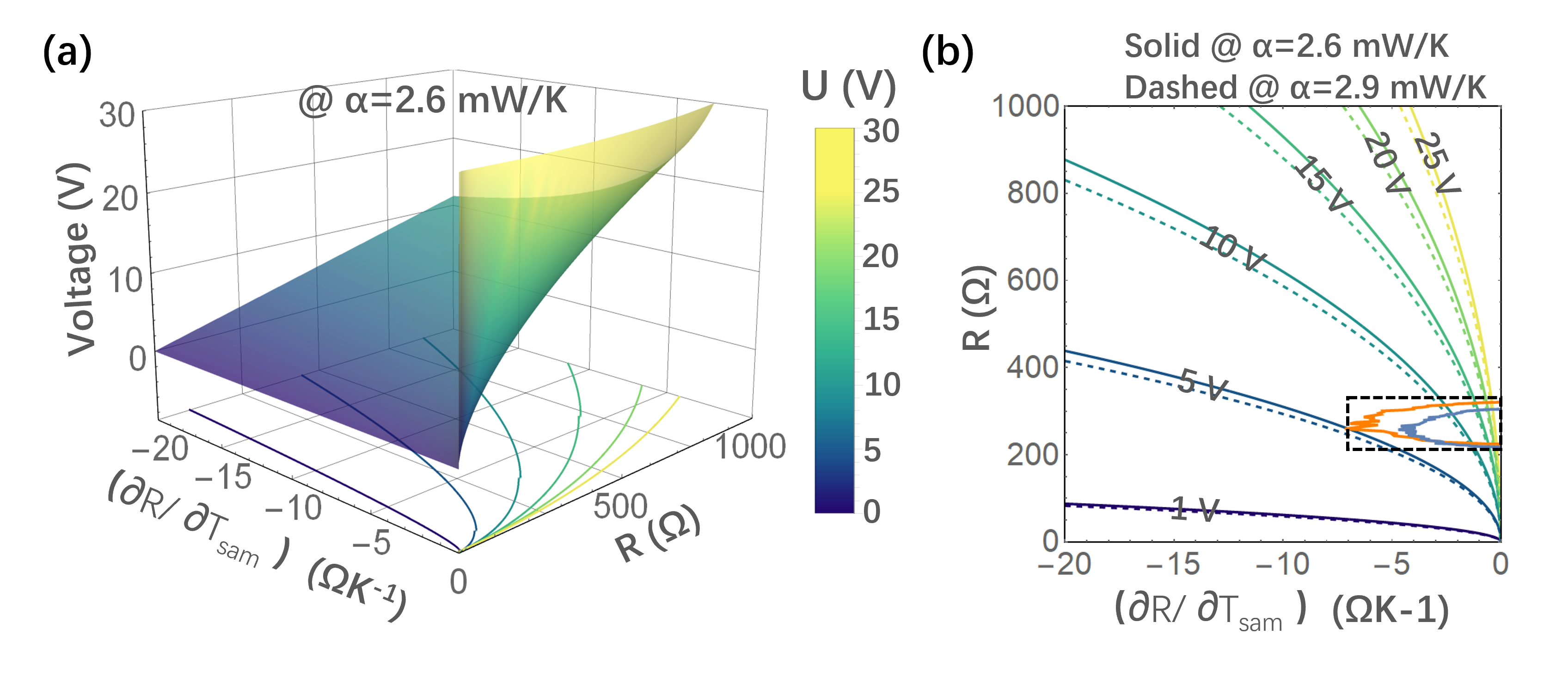}
\caption{The critical condition and the phase diagram for domino-like MPTs. (a) The 3-dimensional surface of critical condition for domino-like MPT and the projection of contours at $T_{env}=308$ K. The color bar is the voltage in unit V. (b) The contours of the surface of the critical condition. The solid (dashed) colored curves are calculated at $T_{env}=308$ K ($T_{env}=360$ K). The orange (blue) curve marked in the black dashed box is $R-\partial R$/$\partial T_{sam}$ curve of the heating (cooling) branch in our experiments.}
\label{6}
\end{figure}

The above results indicate that, although every kinds of driving force have different differential equations, they have an identical critical condition, namely the inequation (4), to occur the domino-like MPT. Thus, the critical condition for domino-like MPTs is independent of the driving force. The inequation (4) constitutes a quadruple phase diagram which consists of four variables: $\partial R/\partial T_{sam}, U, R$ and $\alpha$. Transformed from inequation (4), we obtain the condition of voltage $U\geq\sqrt{-\alpha R^2(\partial R/\partial T_{sam})^{-1}}$ for the occurrence of domino-like MPT. Thus, the function
\begin{equation}
    U=\sqrt{-\alpha R^2(\frac{\partial R}{\partial T_{sam}})^{-1}}
\end{equation}
is a 3-dimensional surface separating the regions of domino-like MPT and conventional MPT. Figure \ref{6} (a) is the 3-dimensional surface calculated at $T_{env}=308$ K ($\alpha$ depends on $T_{env}$). From inequation (4), it is easy to know the region above the 3-dimensional surface will experience the domino-like MPT at $T_{env}\leq308$ K. In figure \ref{6} (b), taking the 10 V contours for example, all of the states in the region below the 10 V solid (dashed) contour will occur domino-like MPT at $T_{env}\leq308$ K ($T_{env}\leq360$ K) under the 10 V voltage. In addition, because $\alpha$ tends to be consistent when $T_{env}>350$ K (see figure \ref{2} (g)), the contours calculated at $T_{env}>350$ K are nearly identical to the dashed colored contours. We obtain the $R-\partial R$/$\partial T_{sam}$ curves of the AFM-FM and FM-AFM transitions from our measuring R-T loop with 0.1 V probe voltage displayed in figure \ref{1} (b), and plot them in figure \ref{6} (b). The $R-\partial R$/$\partial T_{sam}$ curves from our experiments are both above the 5 V solid contour. It indicates that our sample cannot occur domino-like MPT at $T_{env}\geq308$ K when the voltage is lower than 5 V, which corresponds well with all of our measuring results in figure \ref{3}, figure \ref{5} (b), and figure S2 in supporting information. Thus, figure \ref{6} is an accurate phase diagram to determine whether the domino-like MPT can occur.

\section{Conclusion}
In summary, we discover a domino-like FeRh MPT that can be induced by the bias voltage higher than the threshold. Via this kind of MPT, (1) we realize a $10^4$ times enhancement of the maximum transition gradient in the magnetic field, which reduces the transition span of the magnetic field from about $6\times10^4$ Oe to lower than $2\times10^3$ Oe; (2) we display a voltage-pulse-triggered nonvolatile switching between pure magnetic phases at room temperature. The driving force has three types: environment temperature, voltage, and the rest of the other including magnetic field, strain et al. The experiment measurements on these three conditions can be well described by three differential equations derived from thermal equilibrium, respectively, which demonstrates the narrowed transition span and domino-like MPT result from the auto adjustment of the Joule heating under a constant voltage during FeRh resistance changing MPT. The occurrence of the domino-like MPT is independent of driving forces. At the end, we show a phase diagram for the domino-like MPTs in theory. These theoretical studies about domino-like MPT also correspond well with our experiments.

This voltage-controlled phase transition span is universally applicable to all phase transition systems with resistance change, especially holds great significance in the field of magnetic-field-driven phase transition. It makes magnetic-field-driven phase transition as easy as magnetic-field-driven magnetization switching. Furthermore, because the spin-transfer torque and spin-orbit torque also act as an effective magnetic field, our work may further lead to effectively switching magnetic phases using spintronic approaches that have already been successfully used for magnetization reversal, which will greatly promote the practical applications of the MPT-based devices.

\section{Experimental section}
\subsection{Sample fabrication.}
We prepare 50 nm FeRh films on the LiNbO${_3}$ single crystal substrate by co-sputtering at room temperature and anneal them at 973 K for 1 hour in a vacuum furnace to form an ordered CsCl-type structure. Then the FeRh thin films are fabricated to be a strip with 1000 $\mu$m long and 100 $\mu$m width using laser direct writing and ion beam etching.
\subsection{R-T loops and R-V loops measurements.}
R-T loops and R-V loops are measured using a HP 6552A DC power supply, a KEITHLEY 2700, and a home-made heating equipment. The bias voltages are applied by a HP 6552A DC power supply. The resistance of FeRh is obtained via dividing the bias voltage by the current. In the measurements of R-V loops, we tune the environment temperatures to adjust the MPT occurring near various certain voltages.
\subsection{R-H loops measurements.}
The measurement of the R-H loop in figure \ref{5} (a) is carried out with a PPMS equipment, at 350 K and using a 0.1 mA probe current. Other R-H loops are measured using an electromagnet, a HP 6552A DC power supply, and a KEITHLEY 2700. In the measurements of figure \ref{5} (b), in order to shift the AFM-FM transition under various bias voltages to the magnetic field between 0 and 15 kOe, we adjust the environment temperature using the home-made heating equipment to shift the transition point.
\subsection{Magnetic phase domain imaging.}
AFM-FeRh and FM-FeRh have different optical reflectivity, thus we can use this property to image the FeRh magnetic phase domain. The observations of magnetic-field-induced MPT are performed using a MOKE microscopy (Evico Magnetics). The observation region is in the blue dashed rectangle in figure \ref{1} (a). For imaging the AFM-FM MPT, we set the state at zero magnetic field as the background image and observe the difference between the live and background images when the magnetic field increased. Before imaging the FM-AFM transition, we keep the magnetic field at 2000 Oe and set the voltage to 8 V which is enough to induce FeRh to a complete FM state, then the voltage is decreased to 7.00 V. We set this state at 2000 Oe and 7 V voltage as the background image, and observe the difference between the live and background images when the magnetic field decreases.

\section{Data availability}
The data supporting these findings are available from the corresponding author on request.

\section{Author contributions}
H.L.W. conceived the idea. J.B.W. and Q.F.L supervised the project. H.L.W., Q.F.Z and X.Q.W. prepared the samples. H.L.W., C.B.Z. and J.T.X carried out the measurements. H.L.W., Q.F.L., C.B.Z., S.F.Z. and J.W.W. discussed the data. H.L.W. derived the theory and performed the calculations. H.L.W. and Q.F.L. wrote the paper.

\section{Competing interests}
The authors declare no competing interests.

\section{Acknowledgement}
This work is supported by the National Natural Science Fund of China (Grants No. 12174166, 12074158, 12304144, 12104197 and 52201290) and the Fundamental Research Funds for the Central Universities (lzujbky-2024-22).

\section{Supporting information}
Supporting information Available: R-T loops under various voltages in FeRh strip on MgO substrate; Additional calculated MPT loops compared with measured loops; Magnetic-field-induced FeRh FM-AFM transition under the voltage

\section{keywords}
the first-order magnetic phase transition, transition span, domino, thermal equilibrium, magnetic-field-driven magnetic phase transition

\bibliography{paper}

\end{document}